\documentstyle[12pt]{article}

\catcode`\@=11
\long\def\@makefntext#1{
\protect\noindent \hbox to 3.2pt {\hskip-.9pt
$^{{\ninerm\@thefnmark}}$\hfil}#1\hfill}

\def\@makefnmark{\hbox to 0pt{$^{\@thefnmark}$\hss}}  

\def\ps@myheadings{\let\@mkboth\@gobbletwo
\def\@oddhead{\hbox{}
\rightmark\hfil\ninerm\thepage}
\def\@oddfoot{}\def\@evenhead{\ninerm\thepage\hfil
\leftmark\hbox{}}\def\@evenfoot{}
\def\sectionmark##1{}\def\subsectionmark##1{}}

\renewcommand{\thefootnote}{\fnsymbol{footnote}}

\newcounter{sectionc}
\newcounter{subsectionc}\newcounter{subsubsectionc}
\renewcommand{\section}[1]
{\vspace*{0.6cm}\addtocounter{sectionc}{1}
\setcounter{subsectionc}{0}
\setcounter{subsubsectionc}{0}\noindent
        {\normalsize\bf\thesectionc. #1}\par\vspace*{0.4cm}}
\renewcommand{\subsection}[1]
{\vspace*{0.6cm}\addtocounter{subsectionc}{1}
        \setcounter{subsubsectionc}{0}\noindent
        {\normalsize\it\thesectionc.
        \thesubsectionc. #1}\par\vspace*{0.4cm}}
\renewcommand{\subsubsection}[1]
{\vspace*{0.6cm}\addtocounter{subsubsectionc}{1}
        \noindent {\normalsize\rm
        \thesectionc.\thesubsectionc.\thesubsubsectionc
{}.
        #1}\par\vspace*{0.4cm}}

\newcounter{appendixc}
\newcounter{subappendixc}[appendixc]
\newcounter{subsubappendixc}[subappendixc]

\renewcommand{\appendix}[1] {\vspace*{0.6cm}
        \refstepcounter{appendixc}
        \setcounter{figure}{0}
        \setcounter{table}{0}
        \setcounter{equation}{0}
        \renewcommand{\thefigure}
        {\Alph{appendixc}.\arabic{figure}}
        \renewcommand{\thetable}{\Alph{appendixc}.\arabic{table}}
        \renewcommand{\theappendixc}{\Alph{appendixc}}
        \renewcommand{\theequation}
        {\Alph{appendixc}.\arabic{equation}}
        \noindent{\bf Appendix
        \theappendixc #1}\par\vspace*{0.4cm}}

\def\abstracts#1{{
        \centering
{\begin{minipage}{12.2truecm}
\footnotesize\baselineskip=12pt\noindent
        \centerline{\footnotesize ABSTRACT}\vspace*{0.3cm}
        \parindent=0pt #1
        \end{minipage}}\par}}

\renewenvironment{thebibliography}[1]
        {\begin{list}{\arabic{enumi}.}
        {\usecounter{enumi}\setlength{\parsep}{0pt}
\setlength{\leftmargin 1.25cm}{\rightmargin 0pt}
         \setlength{\itemsep}{0pt} \settowidth
        {\labelwidth}{#1.}\sloppy}}{\end{list}}

\newcounter{itemlistc}
\newcounter{romanlistc}
\newcounter{alphlistc}
\newcounter{arabiclistc}

\newcommand{\fcaption}[1]{
        \refstepcounter{figure}
        \setbox\@tempboxa = \hbox{\footnotesize Fig.~\thefigure. #1}
        \ifdim \wd\@tempboxa > 6in
           {\begin{center}
        \parbox{6in}{\footnotesize
        \baselineskip=12pt Fig.~\thefigure. #1}
            \end{center}}
        \else
             {\begin{center}
             {\footnotesize Fig.~\thefigure. #1}
              \end{center}}
        \fi}

\newcommand{\tcaption}[1]{
        \refstepcounter{table}
        \setbox\@tempboxa = \hbox
        {\footnotesize Table~\thetable. #1}
        \ifdim \wd\@tempboxa > 6in
           {\begin{center}
        \parbox{6in}{\footnotesize
        \baselineskip=12pt Table~\thetable. #1}
            \end{center}}
        \else
             {\begin{center}
             {\footnotesize Table~\thetable. #1}
              \end{center}}
        \fi}

\def\@citex[#1]#2{\if@filesw\immediate\write\@auxout
        {\string\citation{#2}}\fi
\def\@citea{}\@cite{\@for\@citeb:=#2\do
        {\@citea\def\@citea{,}\@ifundefined
        {b@\@citeb}{{\bf ?}\@warning
        {Citation `\@citeb' on page \thepage \space undefined}}
        {\csname b@\@citeb\endcsname}}}{#1}}

\newif\if@cghi
\def\cite{\@cghitrue\@ifnextchar [{\@tempswatrue
        \@citex}{\@tempswafalse\@citex[]}}
\def\citelow{\@cghifalse\@ifnextchar [{\@tempswatrue
        \@citex}{\@tempswafalse\@citex[]}}
\def\@cite#1#2{{$\null^{#1}$\if@tempswa\typeout
        {IJCGA warning: optional citation argument
        ignored: `#2'} \fi}}

 1
 1
 1

\font\ninerm=cmr9

\begin{document}

\newcommand{\st}{\scriptstyle}
\newcommand{\sst}{\scriptscriptstyle}
\newcommand{\mco}{\multicolumn}
\newcommand{\epp}{\epsilon^{\prime}}
\newcommand{\vep}{\varepsilon}
\newcommand{\ra}{\rightarrow}
\newcommand{\ppg}{\pi^+\pi^-\gamma}
\newcommand{\vp}{{\bf p}}
\newcommand{\ko}{K^0}
\newcommand{\kb}{\bar{K^0}}
\newcommand{\al}{\alpha}
\newcommand{\ab}{\bar{\alpha}}
\def\be{\begin{equation}}
\def\ee{\end{equation}}
\def\bea{\begin{eqnarray}}
\def\eea{\end{eqnarray}}
\def\CPbar{\hbox
{{\rm CP}\hskip-1.80em{/}}}

\def\coeff#1#2{{\textstyle{#1\over#2}}}
\def\frac#1#2{{#1\over#2}}
\def\hf{\coeff12}
\def\lr{\leftrightarrow}
\def\rtdelta{\sqrt{\Delta}}
\def\Ord{{\cal O}}
\def\rg{r_\Gamma}
\def\cg{c_\Gamma}
\def\disp{{\rm Disp}}
\def\tree{{\rm tree}}
\def\Atree{A^{\rm tree}}
\def\Aloop{A^{\rm 1-loop}}
\def\Re{{\rm Re}\,}
\def\Im{{\rm Im}\,}
\def\as{\alpha_s}
\def\sstw{\sin^2\theta_W}
\def\cstw{\cos^2\theta_W}
\def\ctw{\cos\theta_W}
\def\qb{\bar{q}}
\def\xb{\bar{x}}
\def\costetn{\langle\cos\theta_n\rangle}
\def\Pe{P_e}
\def\Li{\mathop{\rm Li}\nolimits}
\def\ycut{y_{\rm cut}}
\def\fkss{{\rm FKSS}}
\def\hky{{\rm HKY}}
\def\ew{{\rm EW}}
\def\v{V}
\def\f{F}
\noindent
\hfill {SLAC--PUB--6785}\break
\rightline{March, 1995}
\rightline{(T/E)}
\rightline{   }

\centerline{\normalsize\bf TRIPLE-PRODUCT
SPIN-MOMENTUM CORRELATIONS}
\baselineskip=15pt
\centerline{\normalsize\bf
 IN POLARIZED Z DECAYS TO THREE
JETS\footnote{
Research supported by the US Department of
Energy under
grant DE-AC03-76SF00515 and by the Max Kade Foundation.
Talk given at the 4th International
Conference on Physics Beyond
the Standard Model, Lake Tahoe,
CA, December 13-18, 1994, by Y. S.
} }

\vspace*{0.2cm}
\centerline{\footnotesize ARND BRANDENBURG}
\baselineskip=13pt
\centerline{\footnotesize\it
Institut f\"ur Theoretische Physik,
Physikzentrum,}
\centerline{\footnotesize\it Rheinisch-Westf\"alische
Technische Hochschule Aachen,
52056 Aachen, Germany}

\vspace*{0.2cm}
\centerline{\footnotesize LANCE DIXON and YAEL SHADMI}
\baselineskip=13pt
\centerline{\footnotesize\it Stanford Linear Accelerator Center,
Stanford, CA 94309}

\vspace*{0.8cm} \abstracts{We discuss hard rescattering effects
that
can be measured using CP-even, T$_{\rm N}$-odd triple-product
observables in
polarized $Z$ decays to three jets. We show that the standard model
contributions, from both QCD and electroweak rescattering, are very
small. Thus these measurements are potentially sensitive to
physics beyond the
standard model. We investigate one such contribution which involves
a new gauge boson coupling to baryon number. }

\vspace*{0.8cm}
\normalsize\baselineskip=15pt
\setcounter{footnote}{0}
\renewcommand{\thefootnote}{\alph{footnote}}

In testing the standard model (SM) at higher orders,
or in searching for new physics, one usually has to contend
with the tree-level SM background. Three-jet decays of polarized
$Z$ bosons produced in $e^+e^-$ annihilation
offer the possibility of measuring triple-product correlations
such as\cite{FKSS}
$\langle {\bf S_Z \cdot ({\bf k_1} \times {\bf k_2}) }\rangle$,
where ${\bf k_1}$ and ${\bf k_2}$ are two of the three jet momentum
vectors,
and ${\bf S_Z}$ is the $Z$ polarization
vector, parallel to the beam axis\footnote{%
The $Z$ polarization may be produced either with longitudinally
polarized electron beams,
or via the left-right asymmetry $A_{LR}^{(e)} \approx 16\%$
(``natural'' $Z$ polarization).
}.
Such triple-products are odd under T$_{\rm N}$, which
reverses spatial momenta and spin vectors (but does not
exchange initial and final states) so they arise from either
CP violation or rescattering phases\cite{DKD}.
Here we choose ${\bf k_1}$, ${\bf k_2}$ by energy-ordering the
jets, $E_1 > E_2 > E_3$, so that
$\langle {\bf S_Z \cdot ({\bf k_1} \times {\bf k_2}) } \rangle$
is manifestly CP-even, and therefore only sensitive
to rescattering
phases originating from absorptive parts of loop amplitudes.

The triple-product correlation
$
\langle {\bf S_Z \cdot ({\bf k_1} \times {\bf k_2}) } \rangle
$
 could also
be termed ``event handedness'', by analogy to ``jet handedness''
observables\cite{jethand} in which ${\bf S_Z}$ is replaced by
a jet axis, and ${\bf k_i}$ become momenta of particles inside
a single jet, rather than jet momenta.
Several different
variations of event-handedness observables
can be constructed. Here
we focus on
\begin{equation}
  \langle\cos\theta_n\rangle\ =\ \Bigl\langle
  { {\bf \hat{S}_Z} \cdot ({\bf k_1}\times {\bf k_2 })
    \over |{\bf k_1}\times {\bf k_2}| } \Bigr\rangle\ .
\end{equation}

In a covariant framework, a nonzero $\costetn$
is produced by terms in the
differential cross-section that are proportional to
the Levi-Civita tensor $\varepsilon_{\mu\nu\sigma\rho}$
contracted with four of the five momentum vectors in
$e^+e^- \to qg{\bar q}$.
To contribute to the
cross-section, this contraction must be multiplied by the imaginary
part of some loop integral.
There are several possible sources for imaginary parts in the
standard model.
The dominant one, {\it a priori}, is QCD rescattering (Fig.~1a).
However, as argued below, this contribution vanishes for massless
quarks\cite{mqsupp}, and at the $Z$ pole
is suppressed as $\Ord(m_b^2/M_Z^2)$.
Electroweak rescattering could therefore give comparable effects.

%
%
%
%
%

In the following we discuss the four
relevant SM
contributions to~(1): two types of QCD contributions,
and $W$- and $Z$- exchange loops; as well as one non standard
model
contribution.
For brevity, here we give only some numerical results on the
$Z$ pole.
(Our  analytic results including the $\gamma^*$
contributions will be given elsewhere\cite{us}.)
We do not discuss here long-distance, non-perturbative QCD
effects. These are hard to estimate, although they are probably
suppressed by $\Lambda^2_{QCD}$/$M_Z^2$.

Let us first turn to the $\Ord(m_q^2/M_Z^2)$ vanishing of the QCD
contribution. The loop amplitude can be written as a sum of two
parts:
a divergent part, which is proportional to the tree amplitude,
and a
finite part. The former cannot contribute to~(1), because no
Levi-Civita tensor appears in the interference of the relevant
tree
amplitudes. Thus, the only contribution may come from
imaginary parts
of loop integrals that appear in the finite part. These integrals
depend on dimensionless ratios of kinematic invariants, of the type
$(-s_{ij})$/$(-s_{kl})$ and $M_I^2$/$(-s_{ij})$, where
$s_{ij}=(k_i+k_j)^2$, and $M_I$ is the mass of a particle
propagating
in the loop. In the Euclidean region (all $s_{ij} < 0$),
the integrals
are real. But the process we are considering has all
$s_{ij} > 0$, so
the only ratios that change sign upon going from the
Euclidean to the
physical region are those involving internal masses.
Therefore, the
integrals develop imaginary parts only in the presence
of internal
masses. Since this is a kinematic effect, the contribution
of any loop
amplitude involving an internal mass $M_I$ to the
triple-product
correlations vanishes as $\Ord(M_I^2/M_Z^2)$ for small
$M_I$.

Thus, at the $Z$, the dominant QCD contribution comes
from $b$
rescattering, in diagrams such as Fig.~1a.
This contribution was first calculated by
Fabricius {\it et al.}\cite{FKSS} in the case of virtual
photon exchange
(no axial couplings);
they presented numerical results for two choices of
$m_q$/$\sqrt{s}$.
The contribution has the expected $m_q^2$/$M_Z^2$ suppression
for small
quark mass. A further suppression occurs at the $Z$ due to a
cancellation
between the vector and axial components of the signal.
The obtained contribution (A) to $\costetn$
is shown in Table~1 assuming 100\% $Z$ polarization,
for several values of the three-jet cut,
$y_{ij}\ge y_{cut}$, where
$y_{ij} \equiv\ (k_i+k_j)^2/M_Z^2$.

\begin{table}\begin{center}
\tcaption{SM Contributions to $\costetn$.}
{\centerline{\footnotesize
(For $P_z = P_e = 100\%$, $m_b = 4.5$~GeV, $\alpha = 1/129$,
$\alpha_s = .12$, $\sstw = .23.$) }}
\label{tab:smtab}

\begin{tabular}{c|c|c|c} \hline\hline
$y_{cut}$ & QCD (A) & $W$-exchange & $Z$-exchange \\ \hline
 .08 & $-1.5\times 10^{-6}$ & $-4.3
 \times 10^{-7}$ & $-1.7\times 10^{-7}$ \\
 .04 & $-2.6\times 10^{-6}$ & $-5.1
 \times 10^{-7}$ & $-2.0\times 10^{-7}$ \\
.02 & $-3.4\times 10^{-6}$ & $-4.9
\times 10^{-7}$ & $-2.0\times 10^{-7}$ \\
 \hline\hline
\end{tabular}
\end{center}
\end{table}

The second type of QCD contribution
arises from a massive $b$
quark triangle diagram\cite{HKY} (Fig.~1b), which, because of
Furry's theorem, is proportional to the $Z$-quark axial coupling.
Diagrams with up-type and down-type final-state quarks
generate triple-product correlations of opposite signs and
equal (up to
mass-splittings) magnitudes,
so that here too, only $b$ quark
final states contribute.
This contribution turns out to be 2--3 orders of magnitude smaller
than the one discussed above.

The other possible source for event-handedness in the
SM is an
electroweak loop, where the produced quark pair exchanges a $W$ or a
$Z$ (Fig.~1c). As the latter are massive, no quark mass is required
in
order to get
a non-vanishing effect. Hence, all quark flavors contribute here,
except
that the $b$ quark cannot appear in the final state after $W$
exchange below the $t\bar{t}$ threshold (we neglect off-diagonal
CKM
elements).
The last two columns of Table~1
show typical values of these contributions.

Finally, we investigate the sensitivity of
$\langle\cos\theta_n\rangle$
to a recently proposed\cite{CMBD}
$U(1)$ gauge boson $B$, coupling to baryon number only.
This can be easily done by replacing the $W$ or $Z$ in Fig.~1c
by the new boson. The resulting contribution is
biggest for a $B$ mass
of about 30 GeV, for which, taking the $B$ coupling to be
 $\alpha_B$/9, with
$\alpha_B=0.2$, $\costetn \sim 3\times 10^{-5}$.
Although this can be  an order of
magnitude larger than the SM contribution,
it will still be very difficult to find an event-handedness
signature of this boson at present and future colliders.

\vskip .2 cm

\noindent{\bf Acknowledgements}

\vskip .1 cm

We thank T.~Maruyama and P.~Burrows for suggesting this work,
and J.D.~Bjorken, M.~Peskin and W.~Bernreuther
for useful discussions.


\vskip .2 cm

\noindent{\bf References}



\begin{thebibliography}{9}


\bibitem{FKSS}
K. Fabricius, G. Kramer, G. Schierholz and I. Schmitt,
{\it Phys.\ Rev.\ Lett.}~{\bf 45} (1980) 867.

\bibitem{DKD}
A. De R\'ujula, J.M. Kaplan and E. de Rafael,
{\it Nucl.\ Phys.}~{\bf B35} (1971) 365;
A. De R\'ujula, R. Petronzio and B. Lautrup,
{\it Nucl.\ Phys.}~{\bf B146} (1978) 50.



\bibitem{jethand}
O. Nachtmann, {\it Nucl.\ Phys.}~{\bf B127} (1977) 314;
A.V. Efremov, {\it Sov.\ J.\ Nucl.\ Phys.}~{\bf 28} (1978) 83;
A.V. Efremov, L. Mankiewicz and N.A. T\"{o}rnqvist,
{\it Phys.\ Lett.}~{\bf B284} (1992) 394.

\bibitem{mqsupp}
K. Hagiwara, K. Hikasa and N. Kai,
{\it Phys.\ Rev.}~{\bf  D27} (1983) 84;
K. Fabricius, J.G. K\"{o}rner, G. Kramer,
G. Schierholz and I. Schmitt,
{\it Phys.\ Lett.}~{\bf B94} (1980) 207.

\bibitem{us}
A. Brandenburg, L. Dixon and Y. Shadmi, in preparation.

\bibitem{HKY}
K. Hagiwara, T. Kuruma and Y. Yamada,
{\it Nucl.\ Phys.}~{\bf B358} (1991)
80.


\bibitem{CMBD}
C. Carone and H. Murayama, hep-ph/9411256, hep-ph/9501220;
D. Bailey and S. Davidson, hep-ph/9411355.




\end{thebibliography}
\end{document}